\documentclass[%
 aip,
 amsmath,amssymb,
]{revtex4-2}

\usepackage{dcolumn}
\usepackage[mathlines]{lineno}

\usepackage{afterpage}
\usepackage{longtable}
\usepackage{hyperref}
\usepackage[english]{babel}
\usepackage{dcolumn}
\usepackage{bm}
\usepackage{pifont}
\usepackage{amsfonts}
\usepackage{amsmath}
\usepackage{graphicx}
\usepackage{natbib}
\usepackage{psfrag}
\usepackage{color}
\usepackage{multirow}
\usepackage{cases}
\usepackage{stackengine}
\usepackage{float}
\usepackage{blkarray}
\usepackage{cancel}
\usepackage{mathrsfs}
\usepackage{empheq}
\usepackage[utf8]{inputenc}
\usepackage[T1]{fontenc}
\usepackage{mathptmx}

\newcommand*\patchAmsMathEnvironmentForLineno[1]{%
  \expandafter\let\csname old#1\expandafter\endcsname\csname #1\endcsname
  \expandafter\let\csname oldend#1\expandafter\endcsname\csname end#1\endcsname
  \renewenvironment{#1}%
     {\linenomath\csname old#1\endcsname}%
     {\csname oldend#1\endcsname\endlinenomath}}%
\newcommand*\patchBothAmsMathEnvironmentsForLineno[1]{%
  \patchAmsMathEnvironmentForLineno{#1}%
  \patchAmsMathEnvironmentForLineno{#1*}}%
\AtBeginDocument{%
\patchBothAmsMathEnvironmentsForLineno{equation}%
\patchBothAmsMathEnvironmentsForLineno{align}%
\patchBothAmsMathEnvironmentsForLineno{flalign}%
\patchBothAmsMathEnvironmentsForLineno{alignat}%
\patchBothAmsMathEnvironmentsForLineno{gather}%
\patchBothAmsMathEnvironmentsForLineno{multline}%
}

\providecommand\bnabla{\boldsymbol{\nabla}}
\providecommand\bcdot{\boldsymbol{\cdot}}
\providecommand\upi{\pi}%

\newcommand{\ui}{\mathrm{i}}

\newcommand{\tens}[1]{\mathsf{#1}}

\newcommand{\La}{\mbox{\textit{La}}} 	        

\newcommand{\pd}[2] { \frac{\partial   #1}{\partial #2  } }

\newcommand{\hht}{\skew3\hat{h}}

\newcommand{\Uh}{\skew3\hat{\bm{u}}}
\newcommand{\uxh}{\hat{u}}
\newcommand{\urh}{\hat{v}}
\newcommand{\uth}{\hat{w}}

\newcommand{\ph}{\skew3\hat{p}}

\DeclareMathAlphabet{\mathsfbi}{\encodingdefault}{\sfdefault}{bx}{sl}

\newcommand{\com}[1]{\textcolor{black}{#1}}

\begin{document}

\preprint{Accepted in Phys. Fluids}

\title{Non-axisymmetric modes in ultrathin annular liquid films coating a cylindrical fibre}

\author{D. Moreno-Boza}
\email{damoreno@pa.uc3m.es}
\affiliation{Departamento de Ingenier\'ia T\'ermica y de Fluidos, Universidad Carlos III de Madrid. Avda. de la Universidad 30, 28911, Legan\'es, Madrid, Spain.}
\author{A. Sevilla}
\affiliation{Departamento de Ingenier\'ia T\'ermica y de Fluidos, Universidad Carlos III de Madrid. Avda. de la Universidad 30, 28911, Legan\'es, Madrid, Spain.}

\begin{abstract}
    This paper presents a detailed analysis of the three-dimensional stability properties of an annular liquid film coating a cylindrical fibre in the presence of van der Waals (vdW) interactions, whose influence depends on the wettability of the solid by the liquid. Under wetting conditions, vdW interactions can stabilise a uniform annular film when its thickness is smaller than a critical value that depends only on the fibre radius, the Hamaker constant, and the surface tension coefficient~\citep{quere1990spreading}. In contrast, under non-wetting conditions, both surface tension and vdW forces contribute to destabilise the interface, and non-axisymmetric modes may become dominant depending on the thickness of the film and the relative strength of the surface tension and vdW forces. We perform temporal stability analyses of both the Stokes and lubrication equations of motion, allowing us to reveal the dominant azimuthal mode, as well as the optimal axial wavenumber and the corresponding temporal growth rate, as a function of the relevant governing parameters.
\end{abstract}

\date{\today}
\maketitle


\section{Introduction}
A static annular liquid layer of uniform radius coating the external surface of a cylindrical solid fibre is generically unstable due to the Plateau-Rayleigh mechanism, evolving into a sequence of axisymmetric collars along the fibre which are connected by thin annular lobes. The temporal linear stability of the uniform-radius state was solved by~\citet{goren1962instability} using the full Navier-Stokes equations, leading to an exact dispersion relation for the temporal modes. In subsequent papers~\citep{goren1964shape,roe1975wetting} nonlinear equilibrium solutions under quasi-static conditions were found, finding good agreement with the experimentally recorded interface shapes at different times during the unstable evolution.

Since the linear cut-off wavelength is equal to the circumference of the coating, the typical axial length scales are much larger than the radial ones, thus rendering lubrication theory as the natural framework in developing models able to tackle the linear and non-linear dynamics in a simple and accurate way. Indeed,~\citet{dumbleton1970capillary} showed that the linear stability properties of the film are correctly captured by a leading-order lubrication model, while its nonlinear long-time evolution was studied in a series of more recent papers ~\citep{hammond1983nonlinear,johnson1991nonlinear,yarin1993capillary,lister2006capillary}, which provide a complete picture of the flow for arbitrary times in the absence of van der Waals (vdW) forces. In addition, many studies have been devoted to the annular coating process leading to the type of liquid films considered in the present work~\citep[see][and references therein]{quere1999coating}, and have also considered other effects like the gravitational fall of the film along a vertical fibre, which saturates the Plateau-Rayleigh instability~\citep{frenkel1987annular}, and leads to different non-linear regimes depending on the convective or absolute nature of the instability~\citep{kalliadasis1994drop,duprat2007absolute,ruyer2008modelling,Kalliadasis2011}. Thermocapillary effects~\citep{moctezuma2015azimuthal}, the presence of surfactants~\citep{Carroll1974} and the influence of external electric fields~\citep{wray2013electrostatically} have also been addressed. A closely related configuration is that of a thin annular film coating the inner side of a tube, which has also been widely studied due to its relevance in physiology~\citep[see, for instance,][]{hammond1983nonlinear,Grotberg1994,dietze2015films}. Many more examples can be found in the excellent and thorough reviews due to~\citet{oron1997long} and~\citet{craster2009dynamics}, where annular films are discussed in the more general context of thin liquid films.

When the thickness of the film is smaller than about $100~\textrm{nm}$, long-range van der Waals (vdW) forces become comparable to surface tension forces, and their influence depends crucially on the wettability of the solid by the liquid. In cases where the liquid wets the solid, it was shown by~\citet{quere1990spreading} that vdW interactions are able to stabilise the annular film to a uniform thickness that depends only on the fibre radius, the Hamaker constant associated with the disjoining pressure and the surface tension coefficient of the interface separating the liquid layer from the ambient air. In contrast, if the liquid does not wet the solid the coating is always unstable, since both surface tension and vdW forces contribute to destabilise the interface. The resulting unstable flow then combines the effects of both forces, and it is expected that the film dynamics be modified by vdW interactions when the film thickness is sufficiently small. In particular, the linear stability properties of these annular ultrathin non-wetting films, which to the best of our knowledge have not been reported before in full detail, are of interest to describe the resulting dewetting patterns. In contrast with the case of wetting films, where only axisymmetric modes are unstable, it was concluded by~\cite{lin2002three} using an approximate model that non-axisymmetric modes may also become unstable under non-wetting conditions, and that they may even become dominant depending on the thickness of the film and on the relative strength of surface tension and vdW forces.

Thus, our main objective is to provide a detailed account of the three-dimensional stability properties of annular liquid films in the presence of vdW interactions, including both the wetting and the non-wetting scenarios. The mathematical models are formulated in~\S\ref{sec:formulation}, and the corresponding temporal stability analyses are presented in~\S\ref{sec:LSA} distinguishing between wetting and non-wetting cases. Finally, some conclusions are drawn in~\S\ref{sec:conclusions}.


\section{Formulation of the problem}\label{sec:formulation}

\begin{figure}[t]
    \centering
    \includegraphics[width=\textwidth]{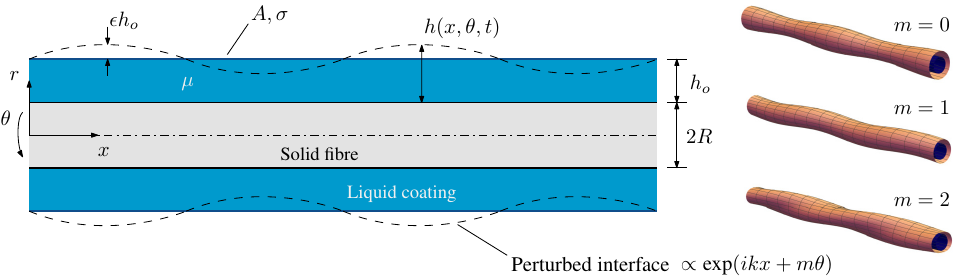}
    \caption{\com{Schematic of the flow configuration illustrating the main governing parameters as well as the cylindrical coordinate system employed in the description. The perturbed air-liquid  interface (dashed lines) is shown for an axisymmetric mode of instability in the main panel, whereas different helical modes $m = 0, 1, 2$ are shown in the right-hand panel. Arbitrary values of $k$ and $\eta$ are used for visualisation.}}
    \label{fig:fig0_sketch}
\end{figure}

An annular film of a Newtonian liquid of density $\rho$ and viscosity $\mu$ coating a cylindrical solid substrate of radius $R$ is considered. The local film thickness will be referred to as $h$ such that the interface has the radial position $r = R + h(x,\theta,t)$ in the cylindrical coordinate system $(x,r,\theta)$. The film, of initial thickness $h_o$, is surrounded by passive atmospheric air at constant pressure $p_a$. The coefficient of surface tension $\sigma$ between the air and the liquid will be taken as a constant, since the presence of temperature gradients and adsorbed surfactants will be disregarded in the present work. The \com{long-range} dispersive van der Waals forces will be modelled by means of a disjoining pressure $\phi^* = A/(6\upi h^3)$, where $A$ is the effective Hamaker constant, which will be considered to be either positive or negative to contemplate, respectively, both the non-wetting and the wetting cases in the analysis. \com{The reader is referred to the review of~\citet{craster2009dynamics} (Section V.B) or the books by~\citet{israelachvili1985} and \citet{Blossey2012} for more details on the derivation of the disjoining pressure. Also, short-range Born repulsion effects will be neglected throughout the analysis. A detailed schematic of the configuration is shown in Figure~\ref{fig:fig0_sketch}.}

\subsection{The Stokes and lubrication equations}\label{subsec:NS}

\com{
The formulation to be presented below will disregard the relative effects of liquid inertia and gravity. On the one hand, the relative importance of inertia may be quantified through the Laplace number $\La=\rho\sigma h_o/\mu^2$, which is a Reynolds number based on the visco-capillary velocity \mbox{$v_{\sigma} = \sigma/\mu$}, and is readily seen to be much smaller than unity in typical configurations of very viscous liquids coating a fibre. On the other hand, gravity effects may also be neglected upon comparing the velocity $v_g = \rho g h_o^2/\mu$, stemming from a balance between viscous and gravitational forces, with the velocity $v_\mathrm{vdW} = |A|/(6 \pi \mu h_o^2)$ arising from a molecular-viscous balance, which yields the rescaled Bond number $v_g/v_\mathrm{vdW} = 6 \pi \rho g h_o^4/|A|$. Therefore, the governing equations of motion under the double limit $La \ll 1$ and $v_g/v_\mathrm{vdW} \ll 1$\footnote{\com{
For a fibre coated by a $1000~\mathrm{cSt}$ silicone oil film of initial thickness $h_o = 100~\mathrm{nm}$, $\rho = 970~\mathrm{kg}/\mathrm{m}^3$, $\mu = 0.97~\mathrm{Pa}\cdot\mathrm{s}$, $\sigma = 27\times 10^{-3} \mathrm{N}/m$ we get $v_g/v_{\mathrm{vdW}} = 1.7\times 10^{-4}$ and $La = 2.8 \times 10^{-6}$.}
} 
are rendered dimensionless upon taking $h_o$, $\sigma/\mu$, $\mu h_o/\sigma$, and $\sigma/h_o$ as the relevant scales for length, velocity, time, and pressure, respectively, and read}
\begin{equation}
\label{eq:ns}
\bnabla \bcdot \bm{u} = 0, \quad \bnabla \bcdot \tens{T} = \bm{0},
\end{equation}
where $\bm{u}=u\bm{e}_x + v\bm{e}_r + w \bm{e}_\theta$ is the liquid velocity field, $\tens{T} = - p\tens{I} + \bnabla \bm{u} + (\bnabla \bm{u})^{\rm{T}} $ is the liquid stress tensor, $p$ is the gauge liquid pressure and $\tens{I}$ is the identity tensor. At the liquid-air interface, $r=\eta+h$, where $\eta=R/h_o$ is the ratio of the cylinder radius to the unperturbed film thickness and $h$ now stands for the dimensionless film thickness, both the kinematic boundary condition,
\begin{equation}
\label{eq:kinematicBC}
\pd{h}{t}=\bm{u}\cdot\bnabla(r-h),
\end{equation} 
and the stress boundary condition,
\begin{equation}
\label{eq:stressBC}
 -\tens{T} \bcdot \bm{n} = \bm{n} (\bnabla \bcdot \bm{n}) + \phi(h) \bm{n},
\end{equation} 
are prescribed, where \com{$\bm{n}=\bnabla(r-\eta-h)/|\bnabla(r-\eta-h)|$ is the outward-pointing normal unit vector, $\phi(h) = \mathcal{A}/h^3$ is the dimensionless disjoining pressure,} and $\mathcal{A} = A/(6 \upi \sigma h_o^2)$ is a dimensionless Hamaker constant that compares the relative importance of vdW and surface tension forces. Note also that $\mathcal{A} = \text{sgn}(A) (h_o/a)^{-2}$, where $a = \sqrt{|A|/(6 \pi \sigma)}$ is the molecular length.\cite{scheludko1962certaines,vrij1966possible,degennes1985wetting}. Finally, no-slip conditions are imposed at the surface of the solid cylinder $r=\eta$, where $\bm{u}=0$.

The lubrication model, derived in Appendix~A, yields the following partial differential equation for the interface position $h(x,\theta,t)$,
\begin{align}
    (\eta + h) \pd{h}{t} + \pd{}{x} \left(G \pd{p}{x} \right) + \pd{}{\theta} \left(H \pd{p}{\theta} \right) = 0, \label{eq:lubmodel}
\end{align} 
where the mobility functions $G=G(\eta,h)$ and $H=H(\eta,h)$, \com{derived in Appendix~\ref{app:derivation_lubrication}}, are given by
\com{
\begin{align}
    & G(h,\eta) =   \frac{h}{16} \left(h+2\eta\right) \left(3 h^2 + 6 h \eta + 2 \eta^2 \right) + \frac{1}{4} (h+\eta)^4 \log \frac{\eta}{h+\eta},  \label{eq:Gx}\\
    & H(h,\eta) =  - \frac{h (h+2\eta)(h^2+2h \eta + 2\eta^2) }{8\eta^2} + \frac{1}{2} (h+\eta)^2 \log \frac{h+\eta}{\eta}, \label{eq:Gtheta} 
\end{align} while the pressure is given by the normal-stress balance 
\begin{equation}
    p = {\mathcal{C}} +  \mathcal{A}{h^{-3}}, \label{eq:pressure}
\end{equation}
being $\mathcal{C}$ twice the mean curvature of the film (see Appendix~\ref{app:derivation_lubrication} for details). The recent study by~\citet{zhao2023slip} on annular thin films coating a fibre derived a similar model but in the absence of molecular forces, $\mathcal{A}=0$, and including the effect of wall slip. The validity of the lubrication model presented above relies in the approximation of the flow being slender, $\epsilon = h_o/\lambda\ll 1$, where $\lambda$ is the characteristic wavelength of the disturbances. We may anticipate that $\lambda \simeq 2\pi h_o/\sqrt{3\mathcal{A}}$ is the dimensional cut-off wavelength~\cite{scheludko1962certaines,vrij1968rupture}. Since the aspect ratio $\eta = R/h_o$ satisfies $\epsilon = h_o/\lambda = h_o R /(R \lambda) = \eta^{-1} R/\lambda \sim \mathcal{A}^{1/2}/\eta^2$, the lubrication approximation is expected to break down for combinations of $\eta$ and $\mathcal{A}$ such that $\epsilon$ is no longer small, i.e. for $\eta \simeq O(1)$ and $\mathcal{A} \lesssim O(1)$, as will be evidenced below.}

Two dimensionless parameters govern the flow at hand, namely the dimensionless Hamaker constant $\mathcal{A}$, and the ratio of the fibre radius to the initial film thickness, $\eta$.


\section{Linear stability analysis}\label{sec:LSA}

\subsection{The unperturbed state}\label{subsec:base}
The base state consists of a static annular film of uniform thickness $h_o$ that depends on the coating process leading to the formation of the film, and may in principle take any value in the context of the stability analysis performed herein. The corresponding dimensional base pressure is given by $p_o=\sigma/(R+h_o)+A/(6\pi h_o^3)$. In the non-wetting case, $A>0$, no stable equilibrium is possible, since both surface tension and vdW forces contribute to destabilise the interface. However, as shown by~\citet{quere1990spreading}, a stable equilibrium is possible in the wetting case, $A<0$, given by a balance between the surface tension and vdW forces, whereby films with thicknesses smaller than $\sqrt{aR}$ are stable.

\subsection{Critical conditions and cut-off wavenumber}\label{subsec:critical}

Following the Plateau-like argument given e.g. by~\citet{scheludko1962certaines}~\citep[see also][]{vrij1966possible,vrij1968rupture}, in which periodic undulations of the free surface with negative extra Gibbs energy tend to grow in time, let us consider the dimensionless excess Gibbs energy $\Delta G$ for our particular configuration in which the initial perturbation to the fibre surface may be written as
\begin{equation}
    r = r_o \equiv  1 + \eta + \delta \cos\left(kx + m\theta\right),
\end{equation} where $\delta \ll 1$ is the amplitude of the undulations. To that end, let $\Delta G = \Delta G_s + \Delta G_\textrm{molec}$, where $\Delta G_s$ is the Gibbs energy associated to the extra surface area  due to the perturbation and $\Delta G_\textrm{molec}$ is the extra interaction energy due to the molecular interactions, which take the form
\begin{align}
    & \Delta G_s = \int_L \int^{2\pi}_0 \left| \bm{X}_\theta \wedge \bm{X}_\varphi \right| \, \mathrm{d}\theta \mathrm{d}x - S_o, \label{eq:DeltaS}\\
    & \Delta G_\textrm{molec} = \int_L \int^{2\pi}_0  \left[ \phi(1) \Delta h + \phi'(1)(\Delta h)^2/2 + \ldots \right] (1+\eta) \, \mathrm{d}\theta \mathrm{d}x, \label{eq:excessG}
\end{align} where $\bm{X} = \left(r_o \cos\theta,r_o \sin\theta, x \right)$ is a parameterization of the corrugated surface, $S_o = 2 \pi (1+\eta) L$ is the unperturbed interface area, $\Delta h = \delta \cos\left(kx+m\theta\right)$, and the axial length $L$ is to be defined below. Note that $\int_h \phi(h)\,\mathrm{d}h$ is the energy of interaction per unit area among the molecules in the film, assumed to be essentially determined by vdW forces. The incompressibility condition
\begin{equation}
    \label{eq:incompressibility_volume}
    \Delta \mathcal{V} = \int_L \int^{2\pi}_0 \int^{r_o}_\eta r \, \mathrm{d}r - V_o= 0,
\end{equation} where $V_o = \pi(1+2\eta)L$ is the unperturbed volume of the liquid fibre, yields the following restriction for the first- and second-order terms
\begin{equation}
    \label{eq:restriction}
    (1+\eta) \delta \int_L \int^{2\pi}_0 \cos (kx+m\theta) \, \mathrm{d}\theta\mathrm{d}x  =
    -\frac{1}{2}\delta^2  \int_L \int^{2\pi}_0 \cos^2(kx+m\theta) \, \mathrm{d}\theta\mathrm{d}x. 
\end{equation} Introducing this restriction onto~\eqref{eq:DeltaS} and letting $L = 2 \upi/k$ be a wavelength produces
\begin{equation}
    \Delta G = \Delta G_s + \Delta G_{\textrm{molec}} =   \frac{ \delta^2 \upi^2}{k (1+\eta) } \left[ {k^2(1+\eta)^2 + m^2 - 1} + (1+\eta)^2 \phi'(1)  \right] + O(\delta^4)
\end{equation}
Following~\citet{scheludko1962certaines,vrij1966possible,vrij1968rupture}, any perturbation is allowed to grow exponentially if the total excess energy $\Delta G$ becomes negative. The cutoff is thus found when $\Delta G = 0$ or, equivalently, to second order,
\begin{equation}
    \label{eq:cutoff_eq}
    k^2(1+\eta)^2 + m^2 - 1 + (1+\eta)^2 \phi'(1)  = 0, 
\end{equation}
for which the cutoff wavenumber $k_c$ is obtained as the positive solution to~\eqref{eq:cutoff_eq}, i.e.,
\begin{equation}
    \label{eq:kc}
    k_c = \sqrt{ -\phi'(1) + \frac{1-m^2}{\left(1+\eta\right)^2} },
\end{equation} where $\phi'(1) = - 3 \mathcal{A}$ for the vdW potential employed throughout this work.
In the wetting case, $\mathcal{A}<0$, only the axisymmetric modes $m=0$ are unstable, and $k_c>0$ for $3|\mathcal{A}|<(1+\eta)^{-2}$. Thus, the film is stable if $3|\mathcal{A}|>(1+\eta)^{-2}$, which corresponds to initial thicknesses smaller than the Qu\'er\'e thickness $h_c$, which is given by the algebraic equation $h_c^2-\sqrt{3} \,a \,(R+h_c)=0$ whose solution is $h_c=3^{1/4}\sqrt{aR}+O(a)$~\citep{quere1990spreading}. It is noteworthy that the same stability condition as in~\citet{quere1990spreading} is obtained, given that their argument is based on the condition that the pressure is an increasing function of the film thickness, but disregards the possibility of small-amplitude harmonic disturbances.

In the non-wetting case, $\mathcal{A}>0$, both axisymmetric and non-axisymmetric modes are potentially unstable. In particular, the axisymmetric and first helical modes, $m=0,\pm 1$ are unstable for any values of $\mathcal{A}$ and $\eta$, while the higher-order azimuthal modes $|m|\geq 2$ are unstable for $\mathcal{A}>(m^2-1)(1+\eta^2)^{-2}/3$. Since different modes compete in a generic low-noise scenario, it is crucial to elucidate the dominant one depending on the values of $\mathcal{A}$ and $\eta$, which requires obtaining the maximum growth rate from the dispersion relation.

\subsection{The dispersion relation}\label{subsec:DR}
The dispersion relation is obtained by introducing~\com{the usual normal-mode decomposition
\begin{equation}
    \label{eq:normalmodedecomp}
    (\bm{u},p,h) = \left(0,\frac{p_o}{\sigma/h_o},1\right) + \delta 
    \left[\Uh(r),\ph(r), \hht\right]
    \mathrm{e}^{\ui (k x + m \theta) + \omega t},
\end{equation}
into the corresponding equations of motion and linearising about the base flow, where $\delta$ is an arbitrarily small quantity, hats are used for eigenfunctions, $\omega$ is the dimensionless growth rate,} $|m| = 0, 1, \ldots$ and $k \in \mathcal{R}$ \com{is the dimensionless wavenumber}. In the case of the Stokes equations~\eqref{eq:ns}, \com{a closed-form solution for the growth rate $\omega = \omega(k,\mathcal{A}, \eta,m)$ can be found analytically in terms of modified Bessel functions. The general process to arrive at such expression is provided in Appendix~\ref{app:stokesDR_detailed}.} In the case of the lubrication model~\eqref{eq:lubmodel}, the solution reads
\begin{equation}
    \omega = \frac{\varphi_1(k,\eta,m) + \varphi_2(k,\eta,m)}{16 \, (1+\eta)}\underbrace{\left[k^2 + \frac{m^2-1}{(1+\eta)^2} - 3\mathcal{A} \right]}_{\textrm{cutoff function}}, \label{eq:DRlub}
\end{equation} where the auxiliary functions 
\begin{align}
    & \varphi_1 = {(2 \eta +1)} \left\{ [2 \eta \, (\eta +3)+3] \, k^2 - \frac{2 m^2}{\eta^2} \, [2 \eta \,  (\eta +1)+1]\right\}, \\
    & \varphi_2 =  8 (1+\eta)^2  \left[ m^2   - \frac{k^2}{2} (1+\eta)^2 \right] \log \frac{1+\eta}{\eta}, 
\end{align} have been defined. A large-$\eta$ expansion for the growth rate~\eqref{eq:DRlub} is available, namely,
\begin{align}
    \omega = \, & k^2 \left(\mathcal{A} -\frac{k^2}{3}\right) 
    +  \frac{k^2 \left(9 \mathcal{A}-40 \, m^2+20\right) +  60 \mathcal{A} \, m^2-3 k^4}{60 \, \eta ^2} + O(\eta^{-3}),\label{eq:DRlub_eta}
\end{align}
which, at leading order, is independent of the azimuthal number $m$ and coincides with the classical result for a planar film~\citep{vrij1966possible}. Three-dimensional effects appear in equation~\eqref{eq:DRlub_eta} in the next-to-leading, $O(\eta^{-2})$ term. \com{Also, in an effort to elucidate the large-$\eta$ behaviour of the dominant mode, given by the pair $(k_M, \omega_M)$, one can find the solution to the condition $\partial_k \omega = 0$ from Eq.~\eqref{eq:DRlub}, wherefrom it is found 
\begin{equation}
    \omega_M = \frac{3 \mathcal{A}^2}{4}  + \frac{9 \mathcal{A}^2+40 \mathcal{A}}{80 \eta ^2}+
    \frac{120 \mathcal{A} m^2
    -21 \mathcal{A}^2-160 \mathcal{A}}{160 \eta ^3}+ O(\eta^{-4}).
\end{equation}
}

\subsection{The dominant azimuthal mode of non-wetting films, $\mathcal{A}>0$}\label{sec:dominant}
The analysis of the Stokes and lubrication dispersion relations reveals that, in both cases, there is only one relevant temporal branch for each combination of $(\mathcal{A},\eta,m)$. 
\begin{figure}
    \centering
    \includegraphics[width=0.95\textwidth]{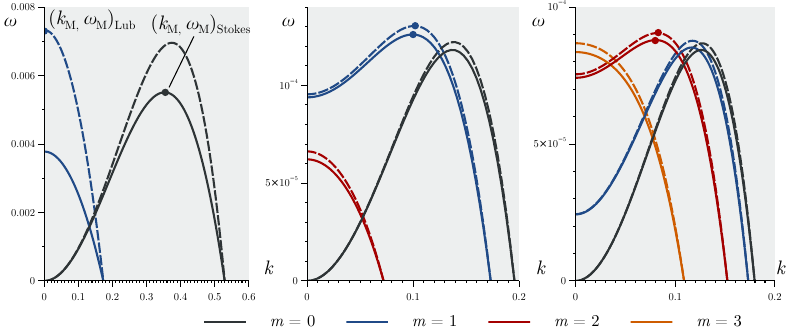}
    \caption{Temporal stability branches for $\mathcal{A} = 0.01$ and $\eta = 1$ (left panel), $\eta = 10$ (central panel), and $\eta = 20$ (right panel), for the azimuthal modes $m=(0,\ldots,3)$. Solid and dashed lines correspond to the Stokes and lubrication results, respectively.}
    \label{fig:1}
\end{figure}

Figure~\ref{fig:1} shows the temporal branches for $\mathcal{A}=0.01$, $\eta=(1,10,20)$ in the right, middle and left panels, respectively, and different values of $m=(0,1,2,3)$. The results obtained using the Stokes dispersion relation are represented with solid lines, while those extracted from the lubrication model~\eqref{eq:DRlub} are plotted with dashed lines. Note first that only for relatively thick films, illustrated by the case $\eta=1$, is there a significant quantitative mismatch between the Stokes equations and lubrication theory, the latter overestimating the growth rate considerably. In contrast, both results are nearly identical for the cases $\eta=(10,20)$, which represent films much thinner than the fibre radius. It is thus deduced that lubrication theory provides a good quantitative description of the linearised dynamics of the flow except for cases where $\eta\lesssim 1$. Secondly, as explained in~\S\ref{subsec:critical}, both the axisymmetric, $m=0$, and first helical, $|m|=1$, modes are unconditionally unstable, while azimuthal modes of increasingly higher order become unstable for increasing values of $\eta$. Note that, according to the Stokes equations, the axisymmetric mode is dominant for $\eta=1$, in that its associated maximum growth rate is larger than that associated to the first helical mode. In contrast, lubrication theory predicts the dominance of the latter mode over the axisymmetric one, as evidenced in the left panel of figure~\ref{fig:1}, thus indicating that lubrication theory must be used with great care when dealing with relatively thick films. In addition, this result suggests that lubrication theory may also fail when applied to the nonlinear flow regimes that take place for large times. However, this issue shall not be clarified in the present contribution, which is focused on the linearised dynamics only. For a relatively thinner film with $\eta=10$, the $|m|=1$ mode is the dominant one according to both the Stokes and lubrication dispersion relations, and the $|m|=2$ mode is also unstable, although with a comparatively smaller maximum growth rate. This trend continues as the film becomes thinner, as can be seen in the left panel, which for $\eta=20$ reveals that the dominant mode is $|m|=2$, and that the $|m|=3$ mode has a growth rate comparable to that of the axisymmetric and $|m|=1$ ones.
\begin{figure}[ht]
    \centering
    \includegraphics[width=0.95\textwidth]{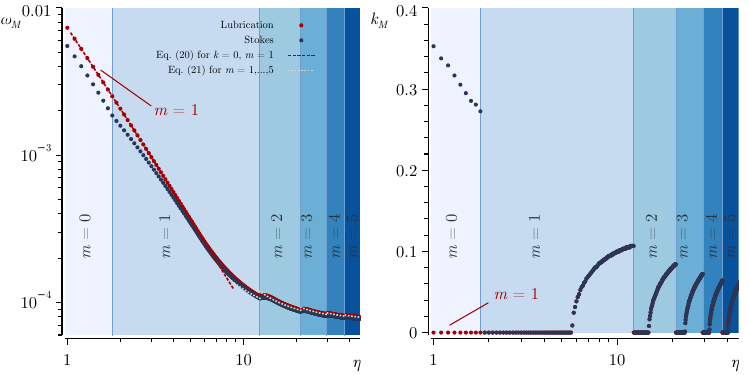}
    \caption{Maximum growth rate $\omega_M$ (left panel) and corresponding wavenumber $k_M$ (right panel) for $\mathcal{A}=0.01$ computed according to the Stokes (blue dots) and lubrication (red dots) dispersion relations. The associated azimuthal mode is indicated explicitly.}
    \label{fig:2}
\end{figure}
\begin{figure}[ht]
    \centering
    \includegraphics[height=0.45\textwidth]{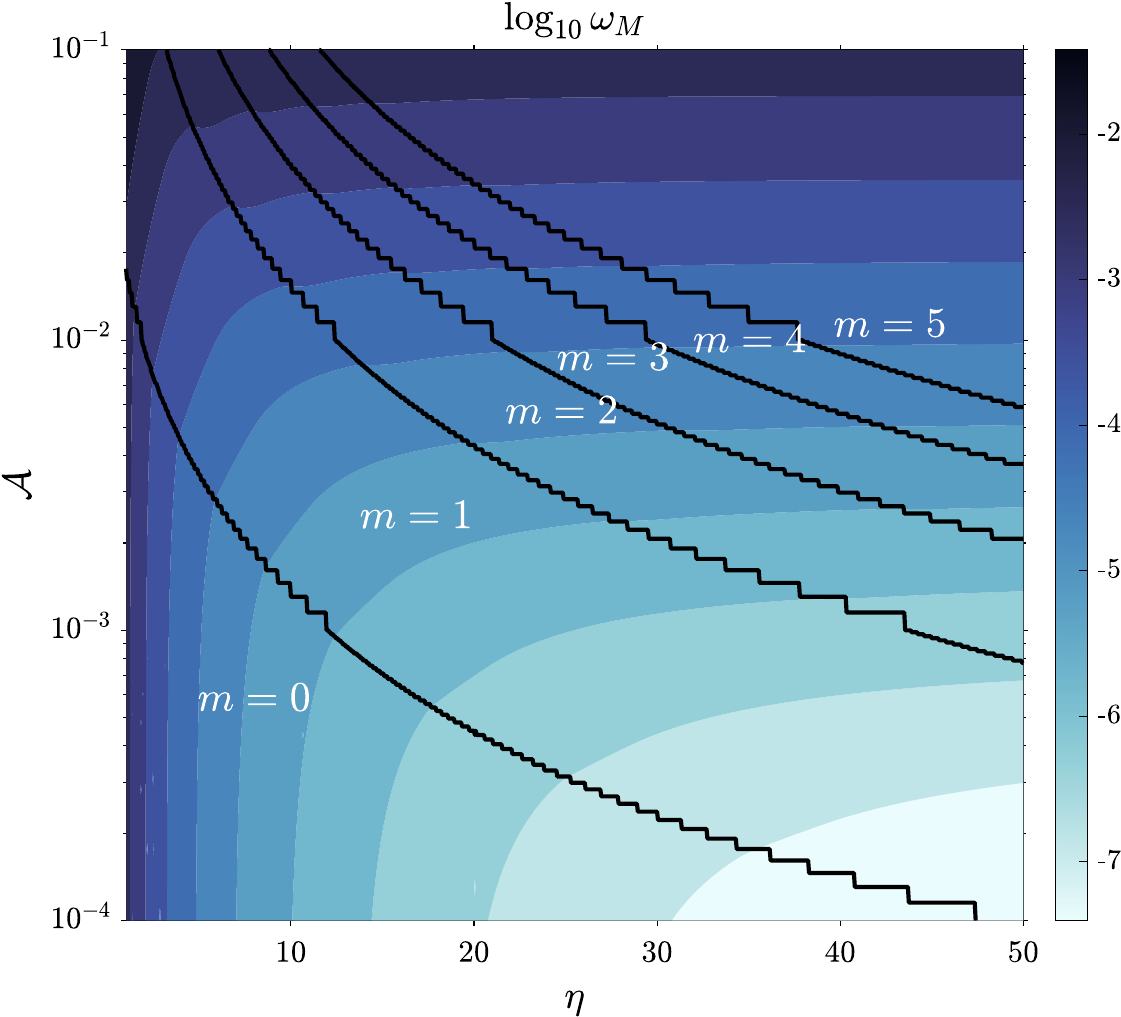}
    \includegraphics[height=0.45\textwidth]{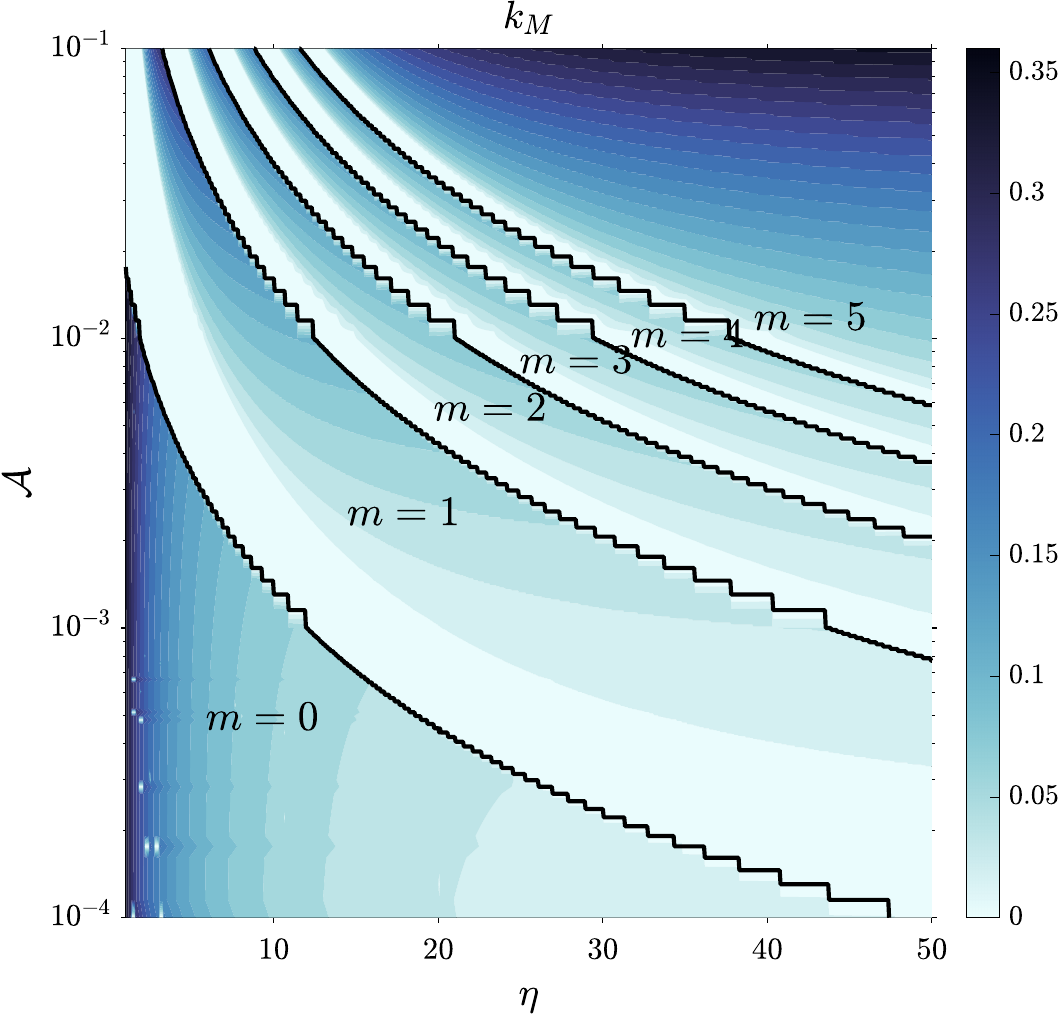}
    \caption{Contour of maximum growth rate (left panel) and corresponding wave number (right panel) over the parametric plane ($\eta$,$\mathcal{A}$) \com{predicted by the linearised Stokes equations}. The boundaries between different dominant azimuthal modes are indicated with solid lines, with the associated values of $m$ indicated explicitly.}
    \label{fig:3}
\end{figure}

Figure~\ref{fig:2}, which represents the maximum growth rate $\omega_M(\eta)$ (left panel) and the associated axial wavenumber $k_M(\eta)$ (right panel) for $\mathcal{A}=0.01$, clearly shows that the order of the dominant azimuthal mode (coloured vertical regions) increases monotonically with $\eta$, starting with $m=0$ for $0<\eta< 1.8$, $|m|=1$ for $1.8<\eta<12.3$, $|m|=2$ for $12.3<\eta<20.9$ and $|m|=4$ for $29.3<\eta<37.4$, the trend continuing for larger values of $\eta$. It is noteworthy that lubrication theory (red dots) fails to capture the transition between the axisymmetric and first azimuthal mode, since it predicts that the dominant mode is always $|m|=1$ for $\eta\lesssim 12$. The parametric behaviour of the optimal wavenumber $k_M$, represented in the right panel of figure~\ref{fig:2}, reveals an interesting non-trivial trend. For values of $\eta<1.8$ the dominant axisymmetric mode has a nonzero associated axial wavenumber with values in the range $0.27<k_M<0.35$. However, when the first azimuthal mode crosses over and becomes dominant, the value of $k_M$ switches to zero in the range $1.8<\eta<5.6$, and becomes strictly positive and increasing as a function of $\eta$ in the remaining range, $5.6<\eta<12.3$. This behaviour of $k_M$ is seen to repeat each time that a transition between the dominant azimuthal modes takes place.

\com{The results of figure~\ref{fig:2} correspond to a fixed value of the dimensionless Hamaker constant, $\mathcal{A}=0.01$. To assess the role played by $\mathcal{A}$, figure~\ref{fig:3} shows contours of $\log_{10}(\omega_M)$ (left panel) and $k_M$ (right panel) in the $(\eta,\mathcal{A})$ parameter plane. The examination of lines of constant $\mathcal{A}$ reveals that the trend associated with the case $\mathcal{A}=0.01$ of figure~\ref{fig:2}, and discussed in the previous paragraph, applies to any value of $\mathcal{A}$, except that the transition to higher-order azimuthal modes takes place for smaller values of $\eta$, i.e. relatively thicker films, when the value of $\mathcal{A}$ increases, i.e. for stronger van der Waals forces relative to surface tension forces. Similarly, the transitions take place for larger values of $\eta$ when the value of $\mathcal{A}$ is decreased. In particular, the boundaries between the different azimuthal modes, represented with solid black lines, have complicated stair-like shapes due to the branch switching behaviour explained in the context of figure~\ref{fig:2}. In addition, the dominant wavenumber is zero in the region surrounding each transition curve, a result that can be understood in the light of the transitions shown in figure~\ref{fig:2} for $\mathcal{A}=0.01$. Indeed, except for the axisymmetric mode $m=0$ which always has a value of $k_M>0$, all the non-axisymmetric modes have an associated value of $k_M=0$ when they become dominant at each transition. It should be noted that the numerical campaign to produce figure~\ref{fig:3} only reaches $m = 5$. Thus, the parametric region corresponding to these higher-order modes is not explored but its trend can be easily conceptualized.}

\section{Concluding remarks}\label{sec:conclusions}
We have presented a detailed analysis of the linear stability of static thin annular films of liquid deposited on cylindrical fibres and subjected to van der Waals interactions, using both the full Stokes equations of motion and a leading-order lubrication description. In the wetting case only axisymmetric modes are unstable, but only when the film thickness is larger than thickness deduced by~\citet{quere1999coating}. In the non-wetting case, our analysis has revealed that both the axisymmetric and the first helical modes are unconditionally unstable, while azimuthal modes of increasing order with $|m|\geq 2$ become unstable when the film thickness is sufficiently smaller than the fibre radius. The transition between dominant modes has been mapped in the parameter plane spanned by the dimensionless strength of the van der Waals forces and the film-to-fibre radius ratio, which are the only dimensionless parameters governing the flow, revealing that the boundaries separating the dominant modes present a surprisingly complicated structure in experimentally relevant parametric ranges.

\com{It is important to note that the maximum growth rates associated to the different azimuthal modes are comparable, especially for large values of $\eta$. For instance, it is deduced from the right panel of figure~\ref{fig:1} that, although the dominant mode is $m=2$ for $\eta=20$ and $\mathcal{A}=0.01$, the values of $\omega_M$ are quite similar for all the unstable modes $m\in \{0,1,2,3\}$. This fact may question the validity of the linear stability analysis in predicting the resulting nonlinear dynamics and dewetting patterns. However, it should be noticed that a small difference in the growth rate of two different normal modes becomes large prior to the nonlinear stages of the flow, provided that the initial amplitude of the noise is small enough. Indeed, let $\omega_M^{(1)}>\omega_M^{(2)}$ be the maximum growth rates of two different modes that share the same initial amplitude $\epsilon$. After a time $t_L$ the relative amplitude is then $\exp[(\omega_M^{(1)}-\omega_M^{(2)})t_L]$, which is of order unity provided that $\omega_M^{(1)}-\omega_M^{(2)} \sim t_L^{-1}$. The key factor for the observation of the dominant mode during the nonlinear regime is then the time $t_L$ during which the linear regime prevails, which can be estimated in terms of the initial amplitude $\epsilon\ll 1$ as $\epsilon \exp(\omega_M^{(1)}t_L)\sim 1 \Rightarrow t_L \sim \log(\epsilon^{-1})/\omega_{M}^{(1)}$. Consequently, the relative amplitude at the end of the linear stage is of order unity provided that $(\omega_M^{(1)}-\omega_M^{(2)})/\omega_M^{(1)} \sim (\log(\epsilon^{-1}))^{-1} \ll 1$. Thus, the smaller the amplitude of the initial noise, the smaller the relative difference between the growth rates of the dominant and subdominant modes that is needed to reach order-unity relative amplitudes at the end of the linear stage.}

\com{As a final remark, our findings indicate the need for further theoretical and experimental explorations. Based on our conclusions, non-axisymmetric modes should be observed in annular dewetting experiments, and to the best of our knowledge, this phenomenon has not been documented in prior studies}. In addition, a study of the nonlinear and large-time dynamics of the film is mandatory, since nontrivial three-dimensional dewetting patterns are expected to arise as a consequence of the non-axisymmetric stability modes reported \com{herein}.

\begin{acknowledgments}
This work has been supported by the Spanish MINECO under project PID2020-115655GB-C22, partly financed through FEDER European funds. The authors thank Dr. A. Mart\'inez-Calvo for fruitful discussions.
\end{acknowledgments}

\vspace{0.5cm}
\noindent\textbf{Declaration of Interests}\\
The authors report no conflict of interest.

\begin{appendix}


\section{Derivation of the leading-order lubrication model}\label{app:derivation_lubrication}

We present here the derivation of the Reynolds equation for the evolution of the annular film, \com{where the main validity assumption relies on the parameter $\epsilon = h_o/\lambda = \mathcal{A}^{1/2}/\eta^2$ being small.} We first recall that the kinematic condition at $r = \eta + h$ reads
\begin{equation}
    \label{eq:lub_kinem}
    \pd{h}{t} + u \pd{h}{x} + \frac{w}{r} \pd{h}{\theta} = v.
\end{equation} Integration of the continuity equation across the liquid layer produces
\begin{equation}
    \int^{\eta + h}_\eta \left( r \pd{u}{x} + \pd{}{r}\left(r v\right) + \pd{w}{\theta} \right) \, \mathrm{d}r = 0,
\end{equation} which upon application of the Leibniz integral rule and use made of~\eqref{eq:lub_kinem} yields
\begin{equation}
    \label{eq:lub_cont}
    (\eta  +  h) \pd{h}{t} + \pd{}{x} \int^{\eta + h}_\eta r \pd{u}{x} \,\mathrm{d}r + \pd{}{\theta}\int^{\eta+h}_\eta \pd{w}{\theta} \,\mathrm{d}r = 0.
\end{equation}
The velocities $u$ and $w$ are found from integration of the momentum equation~\com{ under the assumption of negligible streamwise and azimuthal viscous diffusion, respectively, and negligible radial pressure gradients compared with the axial and azimuthal ones, as a result of the slender approximation. This yields}\footnote{
\com{
A fully detailed asymptotic expansion of the momentum equation and the corresponding boundary conditions is available in the work by~\citet{zhao2023slip}
}
}
\begin{equation}
    \pd{p}{x} = \frac{1}{r} \pd{}{r}\left( r \pd{u}{r} \right), \quad
    \frac{1}{r} \pd{p}{\theta} = \pd{}{r} \left( \frac{1}{r} \pd{}{r} (r w) \right), 
\end{equation} together with the no-slip boundary condition $u = w = 0$ at the cylinder wall $r = \eta$ and vanishing \com{leading-order} tangential stresses $\tau'_{xr} = \partial_r u  = 0$ and $\tau'_{\theta r } = \partial_r w - w/r = 0$ at the free surface $r = \eta + h$, to yield 
\begin{equation}
    u      = \frac{1}{4} \pd{p}{x}      \left[ r^2 - \eta^2 - 2(h+\eta)^2 \log \frac{r}{\eta} \right], \,\,\,  
    w = \frac{1}{4} \pd{p}{\theta} \left[ 2 r \log \frac{r}{\eta} + \frac{(h+\eta)^2 (\eta^2-r^2)}{r \eta^2}  \right].
\end{equation} Inserting these profiles into~\eqref{eq:lub_cont} produces the following evolution equation for $h$
\begin{align}
    \label{eq:lubmodelapp}
    (\eta + h) \pd{h}{t} + \pd{}{x} \left(G \pd{p}{x} \right) + \pd{}{\theta} \left(H \pd{p}{\theta} \right) = 0,
\end{align} where the mobility functions $G$ and $H$ are given by
\begin{align}
    & G(h,\eta) =   \frac{h}{16} \left(h+2\eta\right) \left(3 h^2 + 6 h \eta + 2 \eta^2 \right) + \frac{1}{4} (h+\eta)^4 \log \frac{\eta}{h+\eta},  \label{eq:Gx}\\
    & H(h,\eta) =  - \frac{h (h+2\eta)(h^2+2h \eta + 2\eta^2) }{8\eta^2} + \frac{1}{2} (h+\eta)^2 \log \frac{h+\eta}{\eta}. \label{eq:Gtheta} 
\end{align} The specification of the pressure through the leading-order normal stress balance completes the description, i.e.,
\begin{equation}
    p = {\mathcal{C}} +  \mathcal{A}{h^{-3}}, \label{eq:pressure}
\end{equation} where $\mathcal{C}$ is (twice) the signed curvature of the liquid-air interface,
\begin{equation}
    \mathcal{C} = \bm{\nabla} \cdot \left(\frac{\bm{\nabla} F}{|\bm{\nabla} F|}\right), \quad \textrm{with} \quad F = r - \eta - h(x,\theta,t) = 0,
\end{equation} 
which, expressed in terms of $h(x,\theta,t)$, yields
\begin{equation}
    \mathcal{C} = \, \frac{1}{n (\eta +h) } - \frac{h_{xx} + h_{\theta\theta}/(\eta +h)^2}{n}   + \frac{1}{n^3} \left[  h_x^2 h_{xx} +  \frac{2 h_x h_\theta h_{x\theta}}{(\eta+h)^2}  
    + \frac{h_\theta^2 \left(\eta+h+h_{\theta\theta}\right)}{(\eta+h)^4}\right]
\end{equation}
with $n=\sqrt{1+h_x^2+h_{\theta}^2/(1+\eta)^2}$ and subscripts now indicating partial derivative. 
\com{
    We note that the mobility function $G$ admits the large-$\eta$ expansion     $G/\eta \simeq - h^3 /3 + O(\eta^{-1})$ and that the leading-order curvature reads 
    $ \mathcal{C} \simeq {1}/{\eta} - {h}/{\eta^2} - h_{xx}, $
    for $h \ll \eta$ and $h_x \ll 1$. Thus, the model~\eqref{eq:lubmodelapp} reduces to  
        \begin{equation}
            \pd{h}{t} + \pd{}{x} \left[\frac{h^3}{3} \left( h_{xxx} - \frac{h_x}{\eta^2} - 3  \mathcal{A} h_x h^{-4} \right)  \right] = 0
        \end{equation}
    which coincides with the axisymmetric models presented in~\citet{hammond1983nonlinear} or~\citet{lister2006capillary} for $\mathcal{A} = 0$ up to a rescaling of the $x$-coordinate.
}




\com{
\section{The Stokes dispersion relation}\label{app:stokesDR_detailed}
Introduction of the normal-mode decomposition into the Stokes equations~\eqref{eq:ns} furnishes the linear problem
\begin{align}
    0 & = \ui k \uxh + \frac{1}{r} \pd{}{r}(r\urh) + \frac{\ui m}{r} \uth, \label{eq:lin_cont}\\ 
    \ui k \ph & = \bnabla^2_{k,m} \uxh,  \label{eq:lin_momx}\\ 
    \pd{\ph}{r} & = \bnabla^2_{k,m} \urh - \frac{\urh}{r^2} - \frac{2 \ui m}{r^2} \uth, \label{eq:lin_momr}   \\
    \frac{\ui m}{r}\ph & = \bnabla^2_{k,m}  \uth - \frac{\uth}{r^2} + \frac{2 \ui m}{r^2} \urh, \label{eq:lin_momth}
\end{align}
where $\bnabla^2_{k,m} = \frac{1}{r} \pd{}{r} \left( r \pd{}{r} \right) - (k^2 + m^2/r^2)$ is the normal-mode Laplacian. Since the perturbed velocity is solenoidal, it follows that $\ph$ satisfies the normal-mode Laplace equation
\begin{equation}
    \label{eq:lin_pressure}
    \bnabla^2_{k,m} \ph  = 0,
\end{equation}
so it can be resolved into modified Bessel functions of the form
\begin{equation}
\label{eq:eigpsol}
    \ph  = A_1 I_{m}(kr) + A_2 K_m{(k r)},
\end{equation} where the $A_i$'s are constants of integration, and $I_m$ and $K_m$ are the modified Bessel functions of first and second kind of order $m$, respectively. Consequently, the most general solution to~\eqref{eq:lin_momx} is
\begin{equation}
    \label{eq:eiguxsol}
    \uxh = A_3 I_m(kr) + A_4 K_m(kr) + \frac{\mathrm{i} r}{2} \left[ A_1 I_{m+1} (kr) - A_2 K_{m-1}(kr) \right].
\end{equation}  Upon introduction of~\eqref{eq:eigpsol} into linear combinations of~\eqref{eq:lin_momr} and~\eqref{eq:lin_momth} and leveraging the solenoidality condition~\eqref{eq:lin_cont}, it follows that 
\begin{align}
    \urh = & -\frac{A_1}{2}  \left[\frac{I_{m+1}(k r)}{k}- \frac{r}{2} \left\{ I_{m+2}(k r)  + I_m(k r) \right\} \right] - \nonumber \\
          &  \frac{A_2}{2}  \left[\frac{2  m K_{m+1}(k r) -  K_{m+1}(k r) }{k}- \frac{ r}{2} \left\{ K_{m+2}(k r) + K_m(k r) \right\} \right]- \nonumber \\
          & \mathrm{i} A_3 I_{m+1}(k r)+ \mathrm{i} A_4 K_{m+1}(k r)+\frac{ \mathrm{i} A_5 }{2} \left[ I_{m+1}(k r)-I_{m-1}(k r) \right] + \nonumber \\ 
          & \frac{\mathrm{i} A_6}{2} \left[ K_{m+1}(k r)-K_{m-1}(k r) \right],
\end{align} and
\begin{align}
    \uth = & \frac{\mathrm{i} A_1}{2}  \left[\frac{ I_{m+1}(k r)}{k}-\frac{ r}{2} \left\{ I_{m+2}(k r) -  I_m(k r)\right\} \right]+ \nonumber \\
          & \frac{\mathrm{i} A_2}{2}  \left[\frac{2 m K_{m+1}(k r) -  K_{m+1}(k r)}{k} - \frac{ r}{2} \left\{ K_{m+2}(k r) - K_m(k r)\right\} \right]-  \nonumber \\
          & A_3 I_{m+1}(k r)+A_4 K_{m+1}(k r)+\frac{A_5}{2}  \left[ I_{m-1}(k r)+I_{m+1}(k r) \right]+ \nonumber \\
          & \frac{A_6}{2}  \left[ K_{m-1}(k r)+K_{m+1}(k r) \right].
\end{align} The six constants of integration $A_1,\ldots,A_6$ need to be found upon prescription of the linearised boundary conditions, which in normal-mode form read
\begin{equation}
    \label{eq:eigbcnoslip}
    \uxh = \urh = \uth = 0, \quad \textrm{at} \quad r = \eta, 
\end{equation}
and
\begin{align}
    -& \ph + 2 \urh' + \left(k^2 - \frac{1-m^2}{(1+\eta)^2} - 3\mathcal{A} \right) \hht  = 0 ,  \\
     & \uxh' + \ui k \urh = 0,  \\
     & \uth' - \uth/(1+\eta) + \ui m \urh /(1+\eta)  = 0, \\
     & \omega \hht = \urh, \label{eq:lin_bc_kinem}
\end{align} at $r = 1+\eta$, where primes indicate radial derivatives. Thus, the problem may be recast into a matrix system of the form $\tens{M} \bcdot \bm{A} = 0$, where $\bm{A} = (A_1, \ldots, A_6)^\mathrm{T}$ and $\tens{M}(\omega,k,\mathcal{A},\eta,m)$ is a $6\times 6$ matrix. The entries $M_{ij}$ of the matrix $\tens{M}$ are listed here for completeness:
\begin{align*}
    & M_{11} = \frac{\ui}{2}  \eta  I_{m+1}(k \eta ),  \ 
    M_{12} = -\frac{\ui}{2}  \eta  K_{1-m}(k \eta ), \
    M_{13} = I_m(\eta  k), \
    M_{14} = K_m(\eta  k),  \
    M_{15} = M_{16} = 0, \\
    & M_{21} = \frac{\eta}{4}   I_m(k \eta )+\frac{\eta}{4}   I_{m+2}(k \eta )-\frac{I_{m+1}(k \eta )}{2 k}, \ 
    M_{22} =\frac{\eta}{4}   K_m(k \eta )+\frac{\eta}{4}  K_{m+2}(k \eta )-\frac{(2 m-1) K_{m+1}(k \eta )}{2 k}, \\
    & M_{23} =-i I_{m+1}(k \eta ), \ 
    M_{24} =i K_{m+1}(k \eta ), \\
    & M_{25} =\frac{1}{2} i I_{m+1}(k \eta )-\frac{1}{2} i I_{m-1}(k \eta ), \ 
    M_{26} =\frac{1}{2} i K_{m+1}(k \eta )-\frac{1}{2} i K_{m-1}(k \eta ),
\end{align*} 
\begin{align*}
    & M_{31} = \frac{\ui}{4} \eta  I_m(k \eta )+\frac{i I_{m+1}(k \eta )}{2 k}-\frac{\ui}{4}  \eta  I_{m+2}(k \eta ), \\
    & M_{32} =\frac{\ui}{4}  \eta  K_m(k \eta )+\frac{i (2 m-1) K_{m+1}(k \eta )}{2 k}-\frac{\ui}{4} \eta  K_{m+2}(k \eta ),\\
    & M_{33} = -I_{m+1}(k \eta ), \ M_{34} = K_{m+1}(\eta  k), \\
    & M_{35} = \frac{1}{2} (I_{m-1}(k \eta )+I_{m+1}(k \eta )), \ M_{36} = \frac{1}{2} K_{m-1}(k \eta )+\frac{1}{2} K_{m+1}(k \eta ), 
\end{align*} 
\begin{align*}
    & M_{41} = \frac{1}{4} \left[ \phi_1 + \frac{\phi_2 I_m(\tilde{k}) + \phi_3 I_{m+1}(\tilde{k}) }{k(1+\eta)^2\omega}   \right], \
      M_{42} = \frac{1}{4} \left[ -\phi_1 + \frac{\phi_4 K_m(\tilde{k}) + \phi_5 K_{m+1}(\tilde{k})  }{k(1+\eta)^2\omega}   \right],\\
    & M_{43} = -\ui \left[ k I_m(\tilde{k}) + k I_{m+2}(\tilde{k}) + 
    \frac{\left\{m^2 - 1 - (3 \mathcal{A}-k^2) (\eta +1)^2\right\} I_{m+1}(\tilde{k})}{(\eta +1)^2 \omega } \right],  \\
    & M_{44} = -\ui \left[ k K_m(\tilde{k}) + k K_{m+2}(\tilde{k}) - 
    \frac{\left\{m^2 - 1 - (3A - k^2)(\eta +1)^2 \right\} K_{m+1}(\tilde{k})}{(\eta +1)^2 \omega }
    \right], \\
    & M_{45} = - \frac{\ui m}{k(1+\eta)^3 \omega} \left[ 2 (\eta +1)^2 k \omega  I_{m-1}(\tilde{k}) +  \phi_6  I_m(\tilde{k}) \right] , \\
    & M_{46} = \frac{\ui}{2} \left[ k K_{m-2}(\tilde{k})  -k K_{m+2}(\tilde{k}) + \frac{2 m \left\{m^2 - 1 - (3 \mathcal{A}-k^2) (\eta +1)^2\right)\} K_m(\tilde{k})}{k(1+\eta)^3 \omega}
    \right],
\end{align*} 
\begin{align*}
    & M_{51} = -i \tilde{k} (m+1) I_{m-1}(\tilde{k}) +\ui \left[\tilde{k}^2+2 m (m+1)\right] I_m(\tilde{k}),\\
    & M_{52} = -i \tilde{k} (m-1) K_{m-1}(\tilde{k}) +\ui \left[\tilde{k}^2-(m-2)   m\right] K_m(\tilde{k}), \\
    & M_{53} =  2 \tilde{k} k I_{m-1}(\tilde{k})-3 m k I_m(\tilde{k}), \ M_{54} = -2 \tilde{k} k K_{m-1}(\tilde{k})-3 m k K_m(\tilde{k}) , \\
    & M_{55} = k m I_m(\tilde{k}),  \ M_{56} = -k m K_m(\tilde{k})
\end{align*} 
\begin{align*}
    & M_{61} = \ui (m+1) \left[\tilde{k}\left(\tilde{k}^2+2 m (m+2)\right) I_{m-2}(\tilde{k})-m \left(3 \tilde{k}^2+4 \left(m^2+m-2\right)\right) I_{m-1}(\tilde{k})\right], \\
    & M_{62} = \ui \left[\tilde{k}\left\{\tilde{k}^2 - 2 m \left(m^2-m-2\right)\right\} K_{m-2}(\tilde{k}) -m \left[\tilde{k}^2 (m-3)+4 (m-2) (m-1) (m+1)\right] K_{m-1}(\tilde{k})\right], \\
    & M_{63} =  4 m k\left[\tilde{k}^2+2 \left(m^2-1\right)\right] I_{m-1}(\tilde{k})- k \tilde{k}\left[\tilde{k}^2+4 m (m+1)\right] I_{m-2}(\tilde{k}), \\
    & M_{63} =  - k \tilde{k} \left[\tilde{k}^2+4 m (m+1)\right] K_{m-2}(\tilde{k})-4 m k  \left[\tilde{k}^2+2 \left(m^2-1\right)\right] K_{m-1}(\tilde{k}), \\
    & M_{63} =  k \tilde{k} \left[\tilde{k}^2+2 m (m+1)\right] I_{m-2}(\tilde{k})-2 m  k \left[\tilde{k}^2+2 \left(m^2-1\right)\right] I_{m-1}(\tilde{k}), \\
    & M_{63} =  - k \tilde{k} \left[\tilde{k}^2+2 m (m+1)\right] K_{m-2}(\tilde{k})-2 m  k \left[\tilde{k}^2+2 \left(m^2-1\right)\right] K_{m-1}(\tilde{k}),
\end{align*}
where $\tilde{k} = k(1+\eta)$ and the definition of the following functions has been made for conciseness: 
\begin{align*}
    & \phi_1 = \tilde{k} I_{m-1}(\tilde{k}),  \ \  \phi_2 = 2 \tilde{k} \left(-3 \mathcal{A} (\eta +1)^2+\tilde{k}^2+m^2-(\eta +1) (m+4) \omega -1\right), \\
    & \phi_3 = 6 \mathcal{A} (\eta +1)^2 (m+2) + \tilde{k}^2 [ 3 (\eta +1) \omega -2 (m+2) ]-2 (m+1) (m+2) \left[m-2 (\eta +1) \omega -1\right], \\
    & \phi_4 = 2 \tilde{k} \left[ \tilde{k}^2-3 \mathcal{A} (\eta +1)^2+m^2+(\eta +1) (3 m-4) \omega -1\right], \\
    & \phi_5 = -6 \mathcal{A} (\eta +1)^2 (m-2)+\tilde{k}^2 [3 (\eta +1) \omega +2 m-4]+2 (m-2) (m+1) (-2 (\eta +1) \omega +m-1), \\
    & \phi_6 = m^2 - 1 - (3 \mathcal{A}-k^2) (\eta +1)^2 - 2 (\eta +1) (m+1) \omega ,
\end{align*}
The dispersion relation is finally obtained upon requiring that $\operatorname{det}(\tens{M}) = 0$. The necessary algebra needed to arrive at the final expression for the growth rate $\omega = \omega(k,\mathcal{A},\eta,m)$, which consists in finding the determinant of the matrix $\mathsf{M}$ and solving for $\omega$, is carried out using {\sc Mathematica}~\cite{Mathematica} and, due to its length, it is not amenable for transcription. By means of validation, the eigenproblem \eqref{eq:lin_cont}--\eqref{eq:lin_momth}, \eqref{eq:eigbcnoslip}--\eqref{eq:lin_bc_kinem} is solved separately using a spectral collocation technique. To that end, the physical domain $\eta < r < \eta + 1$ is mapped onto $N$ Chebyshev nodes defined on the domain $-1 < \xi < 1$ by means of the transformation $2r = 1 + 2 \eta - \xi$. The differentiation matrices are obtained following~\citet{trefethen2000spectral} upon discretization of the Chebyshev domain, which eventually are used to recast the eigenproblem into the generalized form $\left(\mathsf{L}-\omega \mathsf{B}\right)\bm{q}= 0$, where $\mathsf{L}$ and $\mathsf{B}$ are the discretized version of their differential counterparts and $\bm{q} = (\Uh, \ph, \hht)^\mathrm{T}$ is the vector of eigenfunctions. The problem is solved using the {\sc Matlab} routine \texttt{eigs}, which is based on the ARPACK~\cite{lehoucq1998arpack} library. Below is a comparison for the results obtained for the sample case $\mathcal{A}=0.01$, $\eta = 10$, $m = 1$, $k = 0.1$, where the reference computation by {\sc Mathematica} is $\omega = 0.0001260252$ and the results provided by the spectral collocation technique are
\begin{center}
\begin{tabular}{rl}
$N$=5,  & 1.261314041e-04 \\
$N$=10, & 1.260252407e-04 \\
$N$=20, & 1.260252407e-04 \\
$N$=50, & 1.260252379e-04 \\
$N$=100,& 1.260252164e-04 
\end{tabular}
\end{center}
showing that even for $N=5$ Chebyshev nodes, the relative error falls below $10^{-3}$.
}
\end{appendix}

\bibliographystyle{plainnat}
\bibliography{biblio_vdw}

\end{document}